# Easy Axis orientation of Ferromagnetic Films as Described by Third Order Perturbed Heisenberg Hamiltonian

P. Samarasekara and Prabhani Rajakaruna

Department of Physics, University of Peradeniya, Peradeniya, Sri Lanka

**ABSTRACT**

The magnetic easy axis orientation was studied by the third order perturbed Heisenberg Hamiltonian. Ferromagnetic CoPt/AlN multilayer thin films with number of layers N=11, 16 and 21synthesized on fused quartz substrates using dc magnetron sputtering technique have been employed as experimental data. According to experimental research performed by some other researchers, easy axis of these fcc structured ferromagnetic films is oriented in the plane of the film above one particular temperature. Average value of out of plane spin component was plotted against temperature in order to determine the spin reorientation temperature. The spin reorientation temperature was highly sensitive to $2^{nd}$ order magnetic anisotropy constant.

## 1. INTRODUCTION

Ferromagnetic thin films are vastly applied in magnetic memory and microwave devices. Annealed thin films of $NdFe_{11}Co_{1-y}Mo_yN$ have been synthesized on polycrystalline $Al_2O_3$ substrates using rf sputtering [1]. Magnetic easy axis oriented $TbCu_7$ type Sm(CoFeCuZr) films have been deposited on polycrystalline $Al_2O_3$ substrates using rf sputtering [2]. Properties of ferromagnetic thin films depend on the composition of target materials, chamber pressure, type of chamber gas, separation between target and substrate, deposition rate and annealing conditions. However, the orientation of magnetic easy axis depends on the orientation of substrate and the substrate (or deposition) temperature. Si doped ZnO thin films have been fabricated on quartz substrates using sol gel method [3]. In addition, $Sn_{1-x}V_xO_2$ films have been grown on si(111) substrates using sol gel method [4].

The easy axis orientation of ferromagnetic films has been explained using second order perturbed modified Heisenberg Hamiltonian by us previously [5, 14]. The Heisenberg Hamiltonian was modified by including stress induced anisotropy, fourth order anisotropy and demagnetization factor. Then the easy axis orientation of Nickel ferrite [6], Barium ferrite [7] and strontium ferrite [8] thin films was explained using unperturbed Heisenberg Hamiltonian by us. The variation of average value of out of plane ($\bar{S}_z$) spin components were plotted in order to determine the spin reorientation temperatures. Hence the temperature, at which $\bar{S}_z$ reaches zero, was determined. According to our previous studies, the spin reorientation temperature depends on the energy parameters of our modified Heisenberg Hamiltonian. However, the spin reorientation temperature was highly sensitive to stress induced anisotropy [6, 7, 8]. The stress induced anisotropy is considerable compared to magnetic anisotropy. The unperturbed, 2$^{nd}$ order perturbed and 3$^{rd}$ order perturbed energy of spinel ferrite was determined by us [13, 16, 17, 18, 19, 20]. According to our previous experimental data, stress induced anisotropy plays a vital role in magnetic thin films [15].

## 2. MODEL

The modified classical Heisenberg Hamiltonian is given by

$$H = -\frac{J}{2}\sum_{m,n}\vec{S}_m \cdot \vec{S}_n + \frac{\omega}{2}\sum_{m\neq n}(\frac{\vec{S}_m \cdot \vec{S}_n}{r_{mn}^3} - \frac{3(\vec{S}_m \cdot \vec{r}_{mn})(\vec{r}_{mn} \cdot \vec{S}_n)}{r_{mn}^5}) - \sum_m D_{\lambda_m}^{(2)}(S_m^z)^2 - \sum_m D_{\lambda_m}^{(4)}(S_m^z)^4$$

$$- \sum_{m,n}\vec{H} \cdot \vec{S}_m - \sum_m K_s Sin2\theta_m$$

The third order perturbed energy of a thick ferromagnetic film can be finally given as [9]

$$E(\theta) = -\frac{J}{2}[NZ_0 + 2(N-1)Z_1] + \{N\Phi_0 + 2(N-1)\Phi_1\}(\frac{\omega}{8} + \frac{3\omega}{8}\cos 2\theta)$$

$$- N(\cos^2\theta D_m^{(2)} + \cos^4\theta D_m^{(4)} + H_{in}\sin\theta + H_{out}\cos\theta + K_s\sin 2\theta)$$

$$- \frac{[-\frac{3\omega}{4}(\Phi_0 + 2\Phi_1) + D_m^{(2)} + 2D_m^{(4)}\cos^2\theta]^2(N-2)\sin^2 2\theta}{2C_{22}}$$

$$- \frac{1}{C_{11}}[-\frac{3\omega}{4}(\Phi_0 + \Phi_1) + D_m^{(2)} + 2D_m^{(4)}\cos^2\theta]^2\sin^2 2\theta$$

$$- \beta_{12}\{\frac{2\alpha_1^2\alpha_2}{C_{22}C_{11}^2} + (\frac{\alpha_2}{C_{22}})^2[\frac{2\alpha_1}{C_{11}} + \frac{2\alpha_2(N-4)}{C_{22}}]\} - \frac{2\alpha_1^3\beta_{11}}{C_{11}^3} - 2(\frac{\alpha_2}{C_{22}})^3\beta_{22}(\frac{N}{2} - 1) \quad (1)$$

where,

$$\beta_{11} = \frac{\omega}{8}\sin 2\theta(4\Phi_0 + \Phi_1) - \frac{4}{3}\cos\theta\sin\theta D_m^{(2)} - 4\cos\theta\sin\theta(\frac{5}{3}\cos^2\theta - \sin^2\theta)D_m^{(4)}$$

$$+ \frac{H_{in}}{6}\cos\theta - \frac{H_{out}}{6}\sin\theta + \frac{4K_s}{3}\cos 2\theta$$

$$\beta_{22} = \frac{\omega}{8}\sin 2\theta(4\Phi_0 + 2\Phi_1) - \frac{4}{3}\cos\theta\sin\theta D_m^{(2)} - 4\cos\theta\sin\theta(\frac{5}{3}\cos^2\theta - \sin^2\theta)D_m^{(4)}$$

$$+ \frac{H_{in}}{6}\cos\theta - \frac{H_{out}}{6}\sin\theta + \frac{4K_s}{3}\cos 2\theta$$

$$\beta_{12} = \frac{3\omega}{8}\sin 2\theta\Phi_1$$

$$C_{11} = JZ_1 - \frac{\omega}{4}\Phi_1(1 + 3\cos 2\theta) - 2(\sin^2\theta - \cos^2\theta)D_m^{(2)}$$

$$+ 4\cos^2\theta(\cos^2\theta - 3\sin^2\theta)D_m^{(4)} + H_{in}\sin\theta + H_{out}\cos\theta + 4K_s\sin 2\theta$$

$$C_{22} = 2JZ_1 - \frac{\omega}{2}\Phi_1(1 + 3\cos 2\theta) - 2(\sin^2\theta - \cos^2\theta)D_m^{(2)}$$

$$+ 4\cos^2\theta(\cos^2\theta - 3\sin^2\theta)D_m^{(4)} + H_{in}\sin\theta + H_{out}\cos\theta + 4K_s\sin 2\theta$$

$$\alpha_1(\theta) = [-\frac{3\omega}{4}(\Phi_0 + \Phi_1) + D_m^{(2)} + 2D_m^{(4)}\cos^2\theta]\sin(2\theta)$$

$$\alpha_2(\theta) = [-\frac{3\omega}{4}(\Phi_0 + 2\Phi_1) + D_m^{(2)} + 2D_m^{(4)}\cos^2\theta]\sin(2\theta)$$

where, N, J, $Z_0$ and $Z_1$, $\Phi_0$ and $\Phi_1$, $\omega$, $\theta$, $D_m^{(2)}$ and $D_m^{(4)}$, $H_{in}$, $H_{out}$, $K_s$ denotes total number of layers in the film, spin exchange interaction, number of nearest spin neighbors, constants arisen from partial summation of dipole interaction, strength of long range dipole interaction, azimuthal angles of spins, second order and fourth order magnetic anisotropy constants, in-plane internal field, out-of-plane internal magnetic field and stress induced anisotropy factor.

After substituting $\alpha_1$, $\alpha_2$, $\beta_{11}$, $\beta_{22}$, $\beta_{12}$, $C_{11}$ and $C_{22}$ in above equation number 1, the total energy can be determined.

Average value of out of plane spin component is given as

$$\bar{S}_z = \frac{\int_0^\pi e^{-\frac{E}{kT}}\cos\theta\, d\theta}{\int_0^\pi e^{-\frac{E}{kT}}\, d\theta} \quad (2)$$

Here k and T are the Boltzmann's constant and the temperature in Kelvin, respectively. Energy given in above equation number 1 was substituted in above equation number 2 to find $\bar{S}_z$.

## 3. RESULTS AND DISCUSSION

Easy axis orientation of sputtered CoPt/AlN multilayer thin films was explained using our model as following. Only the properties of ferromagnetic COPt layers were considered for these explanations. For fcc ferromagnetic thin films, $Z_0=4$, $Z_1=4$, $\Phi_0=9.0336$, $\Phi_1=1.4294$ [10]. The structure of CoPt below annealing temperature of 600 $^0$C is fcc [11]. Figure 1 shows the variation of $\bar{S}_z$ with temperature for CoPt film with N=11 layers. Thickness of this film was given as 4nm [11, 12]. For J=$10^{-30}$ J, $\omega=10^{-35}$ J, $D_m^{(2)}=10^{-27}$ J, $D_m^{(4)}=10^{-25}$ J, $H_{in}=10^{-27}$ Am$^{-1}$, $H_{out}=10^{-30}$ Am$^{-1}$ and $K_s=10^{-28}$ J, the value of $\bar{S}_z$ reaches zero at 484 K. At 484K, $\bar{S}_z$ decreased by 1.14292% of its initial value. So this graph indicates a strong in plane orientation above this particular temperature of 484 K (211 $^0$C).

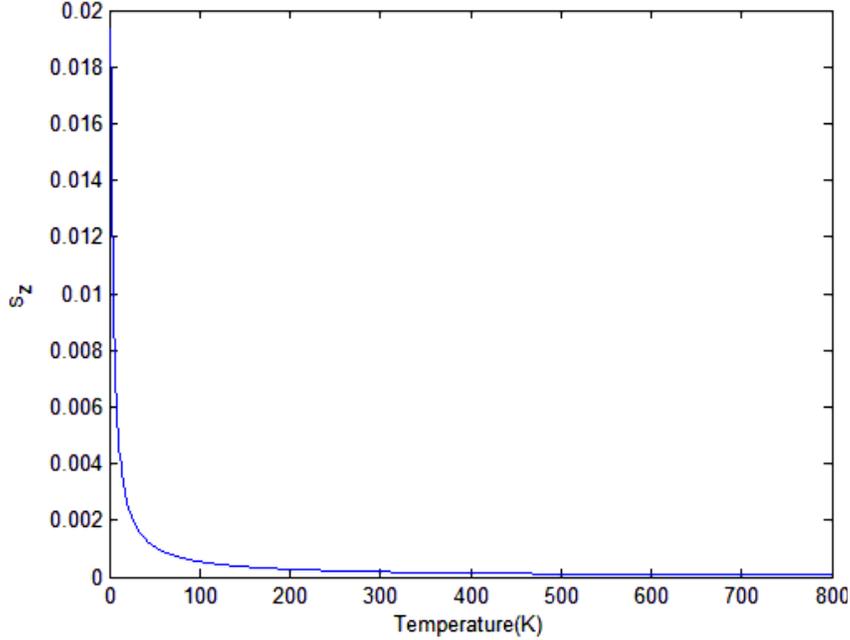

***Fig 1:*** $\bar{S}_z$ *versus temperature for N=11.*

The variation of spin reorientation temperature with energy parameters J, $\omega$, $D_m^{(2)}$, $D_m^{(4)}$, $H_{in}$, $H_{out}$ and $K_s$ was investigated for film with 11 layers. The spin reorientation temperature is highly sensitive to $D_m^{(2)}$. It slightly depends on $\omega$, $H_{out}$ and $K_s$. However, it is not sensitive to J, $D_m^{(4)}$ and $H_{in}$. The dependence of spin reorientation temperature on $D_m^{(2)}$ is shown in figure 2 for $D_m^{(2)}=10^{-26}$ (dashed line) and $10^{-30}$ J (solid line). The spin reorientation temperature for $D_m^{(2)}=10^{-30}$ J is much smaller than that for $D_m^{(2)}=10^{-26}$ J. This implies that the spin reorientation temperature increases with $D_m^{(2)}$ with in the range of $D_m^{(2)}$ from $10^{-27}$ to $10^{-37}$ J. However, it slightly decreases with $D_m^{(2)}$ with in the range of $D_m^{(2)}$ from $10^{-37}$ to $10^{-57}$ J. The other values of energy parameters were kept at J=$10^{-30}$ J, $\omega=10^{-35}$ J, $D_m^{(4)}=10^{-25}$ J, $H_{in}=10^{-27}$ Am$^{-1}$, $H_{out}=10^{-30}$ Am$^{-1}$ and $K_s=10^{-28}$ J for this simulation.

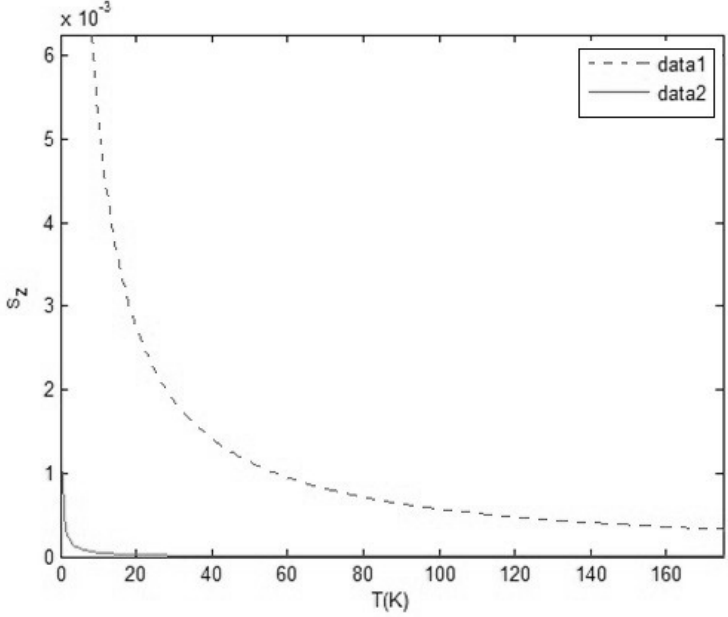

**Fig 2:** $\bar{S}_z$ versus temperature for $D_m^{(2)} = 10^{-26}$ (dshed line) and $10^{-30}$ J (solid line).

The variation of $\bar{S}_z$ with temperature for CoPt film with N=16 layers was also investigated. Thickness of this film was given as 6nm [11, 12]. For J=$10^{-44}$ J, $\omega$=$10^{-54}$ J, $D_m^{(2)}$=$10^{-26}$ J, $D_m^{(4)}$=$10^{-25}$ J, $H_{in}$=$10^{-26}$ Am$^{-1}$, $H_{out}$=$10^{-32}$ Am$^{-1}$ and $K_s$=$10^{-28}$ J, the value of $\bar{S}_z$ reaches zero at 453 K. At 453K, $\bar{S}_z$ decreased by 1.14292% of its initial value. Figure 3 shows the variation of $\bar{S}_z$ with temperature for CoPt film with N=21 layers. Thickness of this film was given as 8nm [11, 12]. For J=$10^{-44}$ J, $\omega$=$10^{-54}$ J, $D_m^{(2)}$=$10^{-26}$ J, $D_m^{(4)}$=$10^{-25}$ J, $H_{in}$=$10^{-26}$ Am$^{-1}$, $H_{out}$=$10^{-32}$ Am$^{-1}$ and $K_s$=$10^{-27}$ J, the value of $\bar{S}_z$ reaches zero at 584 K. At 584K, $\bar{S}_z$ decreased by 1.14292% of its initial value. This implies that the film has a strong in plane orientation above 584 K (311 $^0$C).

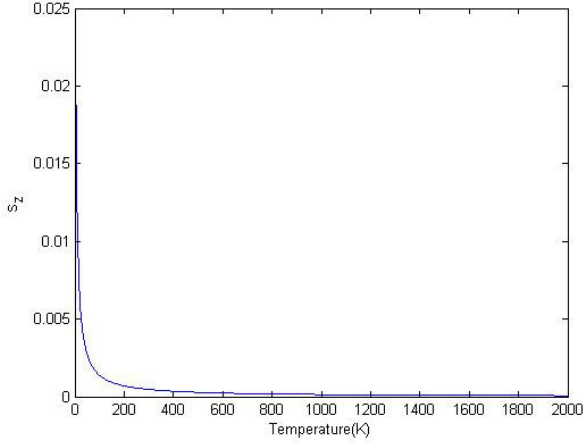

***Fig 3:*** $\bar{S}_z$ *versus temperature for N=21.*

## 4. CONCLUSION

For the first time, the 3$^{rd}$ order perturbed classical Heisenberg Hamiltonian was successfully employed to explain experimental data. According to our theoretical model, the spin reorientation temperature for ferromagnetic films with 11, 16 and 21 layers were 484, 453 and 584 K, respectively. Ferromagnetic films with fcc structure was considered for these simulations. Our theoretical data well agreed with some experimental data published for CoPt/AlN films fabricated on fused quartz substrates by some other researchers [11]. For films with 11, 16 and 21 layers, the average value of out of plane spin component at the spin reorientation temperature decreased to 1.14292% of its initial value. This implied that the easy axis orients in the plane of the film above this particular temperature. The spin reorientation temperature is highly sensitive to second order magnetic anisotropy constant. It is slightly sensitive to ω, $H_{out}$ and $K_s$. However, it is not sensitive to J, $D_m^{(4)}$ and $H_{in}$.